\documentclass[
  reprint,
  superscriptaddress,
  nofootinbib,
  amsmath,amssymb,
  aps,
  prd
]{revtex4-2}

\usepackage{graphicx}
\usepackage{bm}

\usepackage{mathcomp}

\usepackage[colorlinks=true,linkcolor=blue,citecolor=blue,urlcolor=blue]{hyperref}

\begin{document}

\preprint{KUNS-3054, YITP-25-82}

\title{Probing dipolar power asymmetry
with galaxy clustering and intrinsic alignments}

\author{Keita Minato}
\affiliation{Department of Physics, Kyoto University, Kyoto 606-8502, Japan}
\author{Atsushi Taruya}
\affiliation{Center for Gravitational Physics and Quantum Information, Yukawa Institute for Theoretical Physics, Kyoto University, Kyoto 606-8502, Japan}
\affiliation{Kavli Institute for the Physics and Mathematics of the Universe (WPI), The University of Tokyo Institutes for Advanced Study, The University of Tokyo, 5-1-5 Kashiwanoha, Kashiwa, Chiba 277-8583, Japan}
\affiliation{Korea Institute for Advanced Study, 85 Hoegiro, Dongdaemun-gu, Seoul 02455, Republic of Korea}
\author{Teppei Okumura}
\affiliation{Academia Sinica Institute of Astronomy and Astrophysics (ASIAA), No. 1, Section 4, Roosevelt Road, Taipei 10617, Taiwan}
\affiliation{Kavli Institute for the Physics and Mathematics of the Universe (WPI), The University of Tokyo Institutes for Advanced Study, The University of Tokyo, 5-1-5 Kashiwanoha, Kashiwa, Chiba 277-8583, Japan}
\author{Maresuke Shiraishi}
\affiliation{Department of Economics, Management and Information Science, Onomichi City University, Onomichi, Hiroshima 722-8506, Japan}
\affiliation{School of General and Management Studies, Suwa University of Science,  Chino, Nagano 391-0292, Japan}

\date{\today}

\begin{abstract}
We investigate the prospects for probing large-scale power asymmetry through galaxy clustering and intrinsic alignments (IA) in Stage IV galaxy surveys. Specifically, we consider a dipolar modulation in the primordial power spectrum and evaluate the Fisher information matrix using the two-point statistics of both the galaxy clustering and IA. Our analysis reveals that while IA alone provides limited improvement in constraining the modulation amplitude, the cross-spectrum between galaxy density and IA can contribute up to half the constraining power of galaxy clustering, 
especially for surveys with low galaxy bias and high number density of galaxies, such as Euclid.
This demonstrates the potential of IA-clustering cross-correlations as a robust consistency check against systematics, and highlights the complementary roles of galaxy clustering and IA in constraining the asymmetry.
We also show that marginalizing over galaxy bias and IA bias parameters has a negligible impact on the final constraint on the modulation amplitude.

\end{abstract}

\maketitle

\section{Introduction}
\label{sec:intro}
The standard cosmological model, $\Lambda$CDM, assumes that the universe is statistically isotropic and homogeneous on large scales. This assumption is supported by a wide range of cosmological observations. However, recent studies of the cosmic microwave background (CMB), particularly from the Wilkinson Microwave Anisotropy Probe (WMAP)~\cite{Nolta_2004,de_Oliveira_Costa_2004,PhysRevLett.95.071301} and the Planck satellite~\cite{2016,2020_Isotropy}, have revealed several large-scale anomalies that potentially challenge this foundational premise.

Among these anomalies, several features have drawn particular attention as potential imprints of primordial anisotropy in the early universe. Notable examples include the so-called hemispherical asymmetry, a hemispherical variation in the amplitude of CMB temperature fluctuations~\cite{Eriksen_2004,Hansen_2004,Planck_2013_isotropy,2016,2020_Isotropy}, as well as the quadrupole-octupole alignment, which refers to an unexpectedly strong alignment between the CMB quadrupole and octupole moments~\cite{de_Oliveira_Costa_2004}, and the lack of correlations at large angular scales, where the two-point correlation function appears suppressed for angles greater than $60^\circ$ \cite{Hinshaw_1996,Planck_2013_isotropy}. These anomalies, if confirmed, suggest a potential violation of statistical isotropy and may point to new physics beyond the standard inflationary scenario.

Among these anomalies, the hemispherical asymmetry of the CMB temperature fluctuations is often modeled as a dipole modulation, $\Delta T(\hat{n})=\Delta\Bar{T}(\hat{n})\,(1 + A\,\hat{d} \cdot \hat{n})$~\cite{Gordon_2007}, where $\Delta \Bar{T}(\hat{n})$ represents statistically isotropic temperature fluctuations, $A$ is the amplitude of the dipolar modulation, $\hat{n}$ is the line-of-sight direction, and $\hat{d}$ is the preferred direction. 
Latest Planck results indicate 
$\sim 3\sigma$ signal of $A \simeq 0.07$ toward $(\ell,b)\simeq(220\tcdegree,-20\tcdegree)$
in Galactic coordinates \cite{Aiola_2015,2016,2020_Isotropy}.
Such a dipolar modulation motivates alternative early-universe models that go beyond the simplest isotropic inflationary frameworks (see e.g. Refs.\cite{Erickcek:2008sm,Erickcek:2009at,Schmidt:2012ky,Kanno_2013,Lyth:2013vha,Ashoorioon:2015pia,Byrnes:2016uqw}).

While the CMB has been the primary probe of such features, large-scale structure (LSS) provides an independent and complementary window into the early universe (see e.g. Refs.\cite{Hirata:2009ar,Yoon_2014,Appleby_2014,Alonso_2015,Bengaly_2016,Tiwari_2016} for earlier works). Compared to the CMB, LSS probes the matter distribution at lower redshifts and with three-dimensional information, offering the potential for enhanced sensitivity to primordial anisotropies. In particular, dipolar modulation has been studied not only in the context of the CMB (e.g., \cite{Aiola_2015}), but also using LSS probes in more recent analyses. For example, Ref.\cite{Shiraishi_2017} performed a detectability analysis using the three-dimensional galaxy power spectrum, demonstrating that the three-dimensional galaxy distribution provides significantly improved constraints on primordial anisotropy. Ref.\cite{Sugiyama_2017} further applied this framework to actual galaxy survey data, placing tight constraints on 
not dipolar but quadrupolar modulation. These studies highlight the potential of LSS as a promising tool for testing the statistical isotropy of primordial fluctuations, complementary to CMB-based approaches.

Motivated by these developments, in this work we investigate whether additional observables in LSS can enhance sensitivity to such anisotropic signals, focusing specifically on dipolar modulation. Galaxy surveys not only map the spatial distribution of galaxies but also provide information on their intrinsic shapes. These shapes are affected by tidal fields during galaxy formation, leading to a phenomenon known as intrinsic alignment (IA). Although IA has often been considered a contaminant in weak lensing studies \cite{Croft_2000,2004PhRvD..70f3526H}, it actually encodes valuable cosmological information about the large-scale tidal and density fields \cite{Joachimi_2010,Chisari_2013,Chisari_2016,Okumura_2020,Okumura_2020_2,Taruya_2020,Masaki_2020,Akitsu_2021,Akitsu_2021_2,Okumura_2022}.

For star-forming galaxies, the IA signal obtained from conventional galaxy-shape estimators is generally weak. Recent work~\cite{Shi_2021}, however, proposed a fixed-aperture estimator that characterizes the projected light distribution on the scale of the host halo around each ELG. Using mock images from IllustrisTNG, including sky noise and seeing, they obtained a linear-scale response $A_{\rm IA}\simeq(13$--$15) \pm 3.0$ over $0.5\leq z\leq2$, whereas the conventional estimator gave a result consistent with a null detection. The aperture-defined shapes also reproduced the alignment amplitude of the host dark-matter halos. This result motivates treating halo-scale light distributions around spectroscopically selected galaxies as effective IA observables for future surveys.

In particular, the observed ellipticities are two-dimensional projections on the sky. While this projection limits direct access to the full three-dimensional shape information, it still allows for the extraction of statistical correlations that can reveal the underlying anisotropies. Moreover, since the weak lensing contribution is typically small compared to the intrinsic shape dispersion, we neglect lensing distortions in this work and focus solely on the IA signal.

In this paper, we examine the feasibility of testing the dipolar power asymmetry in primordial fluctuations by combining IA with galaxy clustering. Because IA is sensitive to the tidal field at the time of galaxy formation, it retains a memory of the primordial density field. 
By combining intrinsic alignments with galaxy clustering, we extend the analysis 
to include IA statistics and perform a Fisher forecast to assess the potential 
improvement in constraining dipole modulation compared to clustering alone.

As a first step, we perform a Fisher forecast to assess the ability of upcoming galaxy surveys, such as \textit{Euclid} \cite{Euclid2025} and \textit{DESI} \cite{desicollaboration2016desi}, to constrain dipole anisotropy using this approach. Our analysis quantifies the potential improvement when IA is included, and clarifies how LSS data can be used to shed new light on the nature of the large-scale anomalies observed in the CMB.

The rest of this paper is organized as follows.
In Section~\ref{sec:dipole}, we introduce the model of the dipolar power asymmetry considered in this study.  
Section~\ref{sec:probesandstatistics} defines the cosmological probes used in our analysis, specifically, galaxy clustering and IA, and their associated auto- and cross-power spectra, clarifying their relationship to the underlying matter power spectrum.  
In Section~\ref{sec:biposh}, we present the Bipolar Spherical Harmonics (BipoSH) formalism, which provides a powerful framework for quantifying the power asymmetry, and examine how dipolar modulation manifests itself in the expansion coefficients.  
Section~\ref{sec:Fisher_matrix} describes the Fisher matrix analysis based on the BipoSH coefficients, which forms the core of our analysis.  
In Section~\ref{sec:results}, we present the forecast results for the fiducial high-response case $A_{\rm IA}=18$. Section~\ref{sec:conclusion} summarizes our conclusions. Appendix~\ref{Legendre coefficients of power spectra} gives the Legendre coefficients used in the analysis, Appendix~\ref{app:moderate_AIA5} presents the corresponding forecast results for the more moderate value $A_{\rm IA}=5$, and Appendix~\ref{effects_of_marginalizing_over_bias_parameters} discusses the non-degeneracy between the dipolar modulation amplitude and the bias parameters.
\section{Model of Dipolar Modulation}
\label{sec:dipole}
In this paper, we focus on a specific type of power asymmetry in primordial fluctuations: a dipolar modulation.  
Observations of the CMB have revealed hints of a dipole-like, direction-dependent modulation, $\Delta T(\hat{n})=\Delta\Bar{T}(\hat{n})\,(1 + A\,\hat{d} \cdot \hat{n})$ in the CMB temperature fluctuations, 
where $\hat{n}$ denotes the line-of-sight direction.  
In analogy with the CMB temperature fluctuations, the modulation of the matter density fluctuations is modeled as
\begin{align}
    \delta_{\rm m}(\bm{k}, \hat{n}) = \Bar{\delta}_{\rm m} (\bm{k})\left[1+ \sum_M A_{1M}f_{\rm mod}(k)Y_{1M}(\hat{n}) \right],
    \label{dipole_on_density}
\end{align}
where $\Bar{\delta}_{\rm m} (\bm{k})$ represents the isotropic part of the density fluctuations and $Y_{LM}(\hat{n})$ denotes the spherical harmonics with the coefficients $A_{1M}$ corresponding to a dipolar modulation amplitude. 
The coefficients $A_{1M}$ satisfy $A_{1M}^\ast = (-1)^M A_{1,-M}$ and are related to $A$ by $A_{1M} = \tfrac{4\pi}{3}\,A\,Y_{1M}^\ast(\hat{d})$.
Here, the Fourier-space density field explicitly depends on the line-of-sight direction $\hat{n}$, leading to a modulation pattern in the spatial matter distribution\footnote{We consider a collection of finite-volume measurements of the matter or galaxy distribution at distant universe, where the density fields at each patch of the sky can be Fourier-transformed independently of other different patches because of a finite support of density correlations. In this case,  Eq.~\eqref{dipole_on_density} yields a relevant expression to represent the modulated Fourier-space density field. }. 

In this model, $f_{\rm mod}(k)$ characterizes the scale dependence of the modulation.  
If the dipolar modulation is scale-independent ($f_{\rm mod}(k) = 1$), CMB observations have placed a constraint of $A = 0.063^{+0.028}_{-0.030}$~\cite{Aiola_2015}.  
However, constraints from the small-scale ($\ell \gtrsim 60$) CMB data suggest that the modulation amplitude decays with multipole $\ell$ as $A \propto (\ell/60)^{-1/2}$~\cite{Aiola_2015}.  
This decay can be translated into a scale-dependent modulation function, $f_{\rm mod}(k) = (k/k^{\rm c})^{-1/2}$, where $k^{\rm c} = 0.005\,\rm Mpc^{-1}$.  

From the anisotropic matter field, we can define an anisotropic matter power spectrum under the local plane-parallel approximation:
\begin{align}
    \langle \delta_{\rm m}(\bm k,\hat{n})\, \delta_{\rm m}(\bm k^\prime,\hat{n})\rangle = (2\pi)^3 P_{\rm m}(k,\hat{n})\, \delta_{\rm D}^{(3)} (\bm k+\bm k^\prime).
    \label{matter_power}
\end{align}
with $\delta_{\rm D}^{(d)}$ denoting the $d$-dimensional Dirac delta function.

Assuming $|A_{1M} f_{\rm mod}(k)| \ll 1$, consistent with the constraints from CMB observations, the matter power spectrum can be expanded to first order as \cite{Shiraishi_2017}:
\begin{align}
    P_{\rm m}(k,\hat{n}) = \Bar{P}_{\rm m}(k)\left[1+\sum_M 2A_{1M} f_{\rm mod}(k) Y_{1M}(\hat{n})\right],
    \label{Power_Dipole}
\end{align}
where $\Bar{P}_{\rm m}(k)$ is the isotropic matter power spectrum, defined by
\begin{align}
    \langle \Bar{\delta}_{\rm m}(\bm k)\, \Bar{\delta}_{\rm m}(\bm k^\prime)\rangle = (2\pi)^3 \Bar{P}_{\rm m}(k)\, \delta_{\rm D}^{(3)} (\bm k+\bm k^\prime).
\end{align}

\section{Cosmological Probes and Statistics}
\label{sec:probesandstatistics}
In this section we consider anisotropic fields. 
For notational simplicity, however, we omit the directional dependence $\hat{n}$ in Secs.~\ref{GalaxyClustering} and \ref{IntrinsicAlignmentofGalaxy_IA}, unless it is explicitly required.

\subsubsection{Galaxy Clustering}\label{GalaxyClustering}
The most fundamental probe of the large-scale structure of the universe is the spatial distribution of galaxies \cite{Peebles1980}.
In this paper, we consider distant galaxies where the typical angular size of the observed structures is small enough compared to the scale of the dipolar modulation. Under this condition, the observed anisotropies, which arise as an observational effect, can be effectively described using the plane-parallel approximation. Based on this approximation, the Kaiser formula provides a quantitative relation between the observed galaxy distribution $\delta_{\rm g}$ in redshift space and the underlying mass distribution as follows\cite{Kaiser1987,Hamilton_1998,1992ApJ...385L...5H,Okumura_2010}:

\begin{align}
    \delta_{\rm g}(\bm k,\hat{n}) = \left(b_1 + f\mu^2\right) \delta_{\rm m}(\bm k,\hat{n}).
    \label{Kaiser}
\end{align}
The first term in parentheses represents the bias of galaxies, with redshift dependent bias parameter $b_1(z)$.
Second term describes the observational effect called redshift space distortion 
where $f$ is the linear growth factor which refers to the rate at which small density fluctuations (such as those in galaxies or galaxy clusters) grow over time due to the influence of gravity and cosmic expansion, 
and $\mu = \hat{k}\cdot\hat{n}$ is the direction cosine of the wave vector $\bm k$ with respect to the line of sight $\hat{n}$.
The linear growth factor is defined as
\begin{align}
    f(a)=\frac{d \ln{D(a)}}{d\ln{a}}.
\end{align}
with the scale factor $a$ and the linear growth factor $D$, which describes 
the growth of the matter density fluctuation.

\subsubsection{Intrinsic Alignment of Galaxy}\label{IntrinsicAlignmentofGalaxy_IA}

In addition to galaxy clustering, one of the key probes of large-scale structure is the intrinsic alignment (IA) of galaxies.
It is quantified by the two-component ellipticity field $(\gamma_+,\gamma_\times)$
\footnote{In the context of weak lensing, intrinsic ellipticity is often denoted as $\gamma^{\rm I}$ (or $\epsilon$) to distinguish it from the lensing-induced shear. 
In this paper, however, we do not consider lensing effects, and $\gamma$ is understood to represent intrinsic ellipticity.}
, which is defined with the minor-to-major axis ratio $q$ on the celestial sphere:
\begin{align}
    \left(\begin{matrix}
        \gamma_+\\ \gamma_\times
    \end{matrix}\right)(\bm x) 
    = \frac{1-q^2}{1+q^2}\left(\begin{matrix}
        \cos(2\phi_x) \\ \sin(2\phi_x)
    \end{matrix}\right),
    \label{ellipticity_plus_cross}
\end{align}
with the misalignment angle relative to the reference angle $\phi_x$ \cite{Blazek_2015}
It is more convenient to use rotation-invariant quantities called E-/B-modes $(\gamma_{\rm E,B})$, which is well-understood in CMB polarization and weak lensing measurements \cite{Crittenden_2002}.
They are defined in Fourier space as 
$\gamma_{\rm E}(\bm k,\hat{n})+i\gamma_{\rm B}(\bm k,\hat{n}) = {e^{-2i\phi_k}} \left[\gamma_+(\bm k,\hat{n}) + i\gamma_\times(\bm k,\hat{n})\right]$, where $\gamma_{+,\times}(\bm k,\hat{n})$ are the Fourier components of the ellipticity fields and $\phi_k$ is an azimuthal angle of the wave vector projected onto the celestial sphere, measured from the $x$-axis.
Considering the simplest model of IA, called linear alignment (LA) model \cite{Catelan_2001,2004PhRvD..70f3526H,Okumura_2020,Okumura_2019}, in which IA is modeled as being proportional to the gravitational tidal field, the E-/B-modes are given by
\begin{align}
    \gamma_{\rm E}(\bm k,\hat{n}) &= b_{\rm K} \left(1-\mu^2\right) \delta_{\rm m}(\bm k,\hat{n}), \label{ellipticity_Emode}\\
    \gamma_{\rm B}(\bm k,\hat{n}) &= 0,
    \label{ellipticity_Bmode}
\end{align}
where $b_{\rm K}$ is a shape bias, which statistically relates the strength of the alignment of each galaxy image to the tidal force field of LSS.
It is conventionally characterized by introducing a parameter $A_{\rm IA}$ as 
\begin{align}
    b_{\rm K}(z) & = C_1\rho_{\rm crit} A_{\rm IA}\Omega_{\rm m0}/D(z) \nonumber\\
    &= - 0.01344 A_{\rm IA}\Omega_{\rm m0}/D(z),
    \label{bias_and_A_IA}
\end{align}
with the critical density $\rho_{\rm crit}$, the present-day matter density parameter $\Omega_{\rm m0}$, the growth factor $D(z)$, and a normalization factor $C_1$ \cite{Hirata2004}.
For simplicity, we assume that the IA amplitude \( A_{\rm IA} \) is independent of redshift throughout this paper \cite{Kurita_2020}.

Although Eq.~\eqref{ellipticity_plus_cross} is written in terms of the axis ratio of an individual galaxy image, the subsequent large-scale formalism applies more generally to any spin-2 shape estimator whose response to the tidal field is described by Eq.~\eqref{ellipticity_Emode}. Ref.~\cite{Shi_2021} introduced an aperture inertia tensor constructed from pixels with ${\rm S/N} > 3$ within a projected radius $R_{\rm ap}=500\,h^{-1}{\rm kpc}$ around each ELG. Since this scale is comparable to the virial radius of the host halo, the estimator is sensitive not only to the central galaxy but also to satellites and other luminous structures within the halo. Using this estimator, Ref.~\cite{Shi_2021} found $A_{\rm IA}\simeq 15$ for both ELG samples and their host halos. We therefore interpret the high-amplitude case considered below as an effective IA response of such a halo-sensitive observable.

Moreover, recent studies have explored IA observables beyond conventional galaxy-shape measurements. For example, a DESI Y1 analysis found that the orientations of small-scale galaxy multiplets remain aligned with the large-scale tidal field \cite{Lamman_2024}. Also, Ref.~\cite{Ishikawa_2025} reconstructed cluster-halo shapes from projected subhalo distributions and showed that the resulting IA correlations retain cosmological information. These results suggest that tidal-alignment information can remain accessible through halo-scale structure even when individual galaxy shapes are weakly aligned, providing additional motivation for the high-amplitude case considered below.

In the idealized implementation of Ref.~\cite{Shi_2021}, an aperture shape is assigned to every selected galaxy, and we therefore assume $n_{\rm shape}=n_{\rm g}$. In practice, masking, blending, background subtraction, failed shape measurements, and correlations induced by overlapping apertures may reduce the effective number density or modify the shape-noise covariance. Such observational effects are not included in the present forecast and will require calibration with survey data and realistic image simulations.

\subsubsection{Power Spectrum}
Considering the case that an opening angle between the directions towards target galaxies is small and we can neglect the wide-angle effect (plane-parallel limit), we can calculate the auto/cross power spectra of these cosmological fields:
\begin{align}
    &\left<\delta_{\rm g}(\bm k,\hat{n})\delta_{\rm g}(\bm k^\prime,\hat{n})\right> = (2\pi)^3\delta_{\rm D}^{(3)}(\bm k + \bm k^\prime) P^{\rm gg}(\bm k,\hat{n}),
    \label{gg_power_def}\\
    &\left<\delta_{\rm g}(\bm k,\hat{n})\gamma_{\rm E}(\bm k^\prime,\hat{n})\right> = (2\pi)^3\delta_{\rm D}^{(3)}(\bm k + \bm k^\prime) P^{\rm gE}(\bm k,\hat{n}),
    \label{gE_power_def}\\
    &\left<\gamma_{\rm E}(\bm k,\hat{n})\gamma_{\rm E}(\bm k^\prime,\hat{n})\right> = (2\pi)^3\delta_{\rm D}^{(3)}(\bm k + \bm k^\prime) P^{\rm EE}(\bm k,\hat{n}).
    \label{EE_power_def}
\end{align}
Using Eqs.~(\ref{matter_power}), (\ref{Kaiser}) and (\ref{ellipticity_Emode}), 
$P^{\rm XY}$ (${\rm X,Y} \in \rm \{g,E\}$) 
is described as
\begin{align}
    &P^{\rm gg}(\bm k,\hat{n}) = \left(b_1 + f\mu^2\right)^2 P_{\rm m}(\bm k,\hat{n}), 
    \label{gg_power_explicit}\\
    &P^{\rm gE}(\bm k,\hat{n})= P^{\rm Eg}(\bm k,\hat{n})\nonumber\\
    &~~~~~=
    \left(b_1 + f\mu^2\right)b_{\rm K}\left(1-\mu^2\right)P_{\rm m}(\bm k,\hat{n}),
    \label{gE_power_explicit}\\
    &P^{\rm EE}(\bm k,\hat{n}) = b_{\rm K}^2\left(1-\mu^2\right)^2 P_{\rm m}(\bm k,\hat{n}).
    \label{EE_power_explicit}
\end{align}

When the matter field is statistically isotropic, these power spectra are decomposed with Legendre polynomials $\mathcal{L}_\ell$: 
\begin{align}
    \Bar{P}^{\rm XY}(\bm{k},\hat{n}) = \sum_\ell P_\ell^{\rm XY}(k) \mathcal{L}_\ell(\mu),
    \label{power_Legendre_decomposition}
\end{align}
where the explicit form of the Legendre coefficients $P_\ell^{\rm XY}(k)$ is given in Appendix \ref{Legendre coefficients of power spectra}.

\section{BipoSH Decomposition}
\label{sec:biposh}

In a statistically isotropic and homogeneous universe, the power spectrum can be expanded in terms of Legendre polynomials including the effect of redshift space distortion \cite{Peebles1980,Kaiser1987,Hamilton_1998}, as shown in Eq.~(\ref{power_Legendre_decomposition}). In such analyses, the line-of-sight direction is typically fixed and treated as a special axis, and the apparent anisotropy is evaluated relative to this direction.

However, in the presence of a power asymmetry, another special direction characterizing the asymmetry emerges, in addition to the line of sight. In this case, the observed asymmetry must be characterized with respect to both the line-of-sight and the preferred directions. The standard Legendre expansion, which assumes rotational symmetry about the line of sight, becomes insufficient to capture this more complex angular dependence.

To properly describe these anisotropic signatures, a more general basis is required. The Bipolar Spherical Harmonics (BipoSH) \cite{Varshalovich_1988} provide a suitable decomposition, as they encode correlations between different multipoles and capture anisotropies beyond axial symmetry. The BipoSH basis functions are defined as
\begin{align}
    X_{\ell\ell^\prime}^{LM}(\hat{k}, \hat{n}) =& \{Y_{\ell}(\hat{k})\otimes Y_{\ell^\prime}(\hat{n})\}_{LM} \nonumber \\
    =& \sum_{mm^\prime} \mathcal{C}_{\ell m \ell^\prime m^\prime}^{LM} \, Y_{\ell m}(\hat{k}) Y_{\ell^\prime m^\prime}(\hat{n}),
    \label{BipoSH_def}
\end{align}
where $Y_{\ell m}$ are the spherical harmonics, and
\begin{align}
    \mathcal{C}_{\ell_1 m_1 \ell_2 m_2}^{\ell_3 m_3} = (-1)^{\ell_1 - \ell_2 + m_3} \sqrt{2\ell_3 + 1}
    \begin{pmatrix}
        \ell_1 & \ell_2 & \ell_3 \\
        m_1 & m_2 & m_3
    \end{pmatrix}
    \label{Clebsch_Gordan_coefficients}
\end{align}
are the Clebsch–Gordan coefficients, written in terms of the Wigner 3$j$ symbols$\left(
    \begin{matrix}
        \ell_1&\ell_2&\ell_3\\
        m_1&m_2&m_3
    \end{matrix}
    \right)$.

The multipole indices $L$ and $M$ characterize the anisotropic components. 
The case $L=0$ and $M=0$ represents the contribution from the isotropic component, 
for which the BipoSH expansion reduces to the usual Legendre expansion up to a normalization factor. 
On the other hand, modes with $L>0$ describe contributions from anisotropic components. 
In particular, the dipolar anisotropy considered in this paper corresponds to $L=1$, 
as can be seen from Eq.~\eqref{BipoSH_coefficient_explicit}. 
The allowed values of the indices $(\ell,\,\ell')$ are determined by the Clebsch--Gordan coefficients, 
which impose the condition $|M|\leq L$. 
In the dipole case ($L=1$), the specific pairs of $(\ell,\,\ell')$ that contribute 
are listed below Eq.~\eqref{BipoSH_coefficient_explicit}.

The BipoSH decomposition technique has successfully been applied to the number density correlator in the dipolar asymmetry model \cite{Shiraishi_2017} and also the number density, velocity and ellipticity correlators in other-type asymmetry models \cite{Shiraishi_2017,Bartolo:2017sbu,Akitsu:2019avy,Shiraishi:2020pea,Shiraishi:2023zda}. In this paper, following a similar computational procedure, we deal with the number density and ellipticity correlators in the dipolar asymmetry model.

As shown in Eqs.~(\ref{gg_power_def})–(\ref{EE_power_def}), the power spectra become dependent on the line-of-sight direction $\hat{n}$ when a dipolar modulation is present in the density fluctuations. Accordingly, we decompose the power spectra using the BipoSH basis as
\begin{align}
    P^{\rm XY}({\bm k},\hat{n}) = \sum_{\ell\ell^\prime LM} \,_{\rm XY}\pi_{\ell\ell^\prime}^{LM}(k) X_{\ell\ell^\prime}^{LM}(\hat{k}, \hat{n}),
    \label{power_BipoSH_decomposition}
\end{align}
which allows us to capture the full directional dependence with respect to both $\hat{k}$ and $\hat{n}$.

The BipoSH satisfies the following orthogonality relation:
\begin{align}
    \int d^2\hat{k} \int d^2 \hat{n} X_{\ell\ell^\prime}^{LM}(\hat{k}\cdot\hat{n}) X_{\Tilde{\ell}\Tilde{\ell}^\prime}^{\Tilde{L}\Tilde{M}}(\hat{k}\cdot\hat{n})^\ast
    = \delta_{L,\Tilde{L}} \delta_{M,\Tilde{M}} \delta_{\ell,\Tilde{\ell}} \delta_{\ell^\prime,\Tilde{\ell}^\prime}
    \label{orthogonality_of_BipoSH}.
\end{align}
with $\delta_{i,j}$ denoting the Kronecker delta.

By applying this equation, the BipoSH coefficients are obtained as follows:
\begin{align}
    {}_{\rm XY}\pi_{\ell\ell^\prime}^{LM}(k) = P^{\rm XY}_\ell(k)\left[\frac{4\pi}{2\ell + 1} H_{\ell\ell^\prime 0}^{-1}\delta_{\ell,\ell^\prime}\delta_{L,0}\delta_{M,0}\right. \nonumber\\\left.-2\sqrt{\frac{4\pi(2\ell^\prime+1)}{2\ell+1}}H_{\ell\ell^\prime 1}A_{1M}f_{\rm mod}(k)\delta_{L,1}\right],
    \label{BipoSH_coefficient_explicit}
\end{align}
where $H_{\ell_1\ell_2\ell_3} = \left(
    \begin{matrix}
        \ell_1&\ell_2&\ell_3\\
        0&0&0
    \end{matrix}
    \right)$ 
and $P_\ell^{XY}$ is the Legendre coefficient defined in Eq. (\ref{power_Legendre_decomposition}), 
assuming that there is a dipolar modulation represented by Eq.(\ref{Power_Dipole}).
The first term inside the brackets corresponds to the isotropic component,
while the second term arises from the dipole component.

From the property of the Wigner 3$j$ symbol and Legendre coefficients of power spectra, the modulation due to dipolar asymmetry creates a non-vanishing ${}_{\rm XY}\pi_{\ell\ell^\prime}^{1M}$ for $(\ell, \ell^\prime) = (0, 1)$, $(2, 1)$, $(2, 3)$, $(4, 3)$ and $(4, 5)$.

A practical way to estimate the anisotropic power spectrum through the BipoSH decomposition (Eq.(\ref{BipoSH_coefficient_explicit})) from real data is described in Ref.\cite{Sugiyama_2017}.

\section{Fisher Forecast}
\label{sec:Fisher_matrix}

In this paper, we apply the BipoSH decomposition on the power spectra, and to investigate the statistical error of the modulation amplitude, we evaluate the explicit form of the Fisher matrix as follows.
The Fisher matrix for arbitrary parameters $\theta_i$
from the BipoSH coefficients ${}_{\rm XY}\pi_{\ell\ell^\prime}^{LM}$ is defined by
\begin{widetext}
\begin{align}
    F_{ij} 
    &=\sum_{k_1,k_2}^{k_{\rm max}}\sum_{\mathcal{C}_1,\mathcal{C}_2}   
    \frac{\partial \,_{\rm X_1Y_1}\pi_{\ell_1\ell_1^\prime}^{L_1M_1}(k_1)^\ast}{\partial \theta_i^\ast}
    \left\langle\;_{\rm X_1Y_1}\pi_{\ell_1\ell_1^\prime}^{L_1M_1}(k_1)^\ast \;_{\rm X_2Y_2}\pi_{\ell_2\ell_2^\prime}^{L_2M_2}(k_2)\right\rangle_c^{-1} 
\frac{\partial \,_{\rm X_2Y_2}\pi_{\ell_2\ell_2^\prime}^{L_2M_2}(k_2)}{\partial \theta_j}
    \label{Fisher_def},
\end{align}
where $\mathcal{C}_i$ represents a set of subscripts, $\mathcal{C}_i=\{
\ell_i, \ell_i^\prime, L_i, M_i, {\rm X}_i, {\rm Y}_i\}$.
Here, the covariance matrix
\(
\left\langle\,_{\rm X_1Y_1}\pi_{\ell_1\ell_1^\prime}^{L_1M_1}(k_1)^\ast\, 
\,_{\rm X_2Y_2}\pi_{\ell_2\ell_2^\prime}^{L_2M_2}(k_2)\right\rangle_c
\)
is treated as a matrix indexed by a single composite label $ \mathcal{C}_i
$. That is, the BipoSH coefficients \({}_{X_iY_i}\pi_{\ell_i\ell_i^\prime}^{L_iM_i}(k_i)\) are vectorized over these indices and the covariance is considered as a matrix \(\mathrm{Cov}_{\mathcal{C}_1 \mathcal{C}_2}\) acting on these vector components. The inverse in Eq.~(\ref{Fisher_def}) should thus be understood as the matrix inverse of this vectorized covariance structure.
This formulation allows the Fisher matrix to account for cross-correlations between all BipoSH modes and wavenumbers consistently, including those arising from the power asymmetry. 

The quantity $\left\langle{}_{\rm X_1Y_1}\pi_{\ell_1\ell_1^\prime}^{L_1M_1}(k_1)^\ast \,_{\rm X_2Y_2}\pi_{\ell_2\ell_2^\prime}^{L_2M_2}(k_2)\right\rangle_c$ is the covariance matrix of the BipoSH coefficients of the power spectra ${{}_{\rm XY}\pi^{LM}_{\ell\ell^\prime}(k)}$
, which is translated from covariance matrix of $P^{\rm XY}$ using the orthogonality of BipoSH (Eq. (\ref{orthogonality_of_BipoSH})).
$\left\langle P^{\rm X_1Y_1}(\bm k_1, n_1) \,P^{\rm X_2Y_2}(\bm k_2, n_2) \right\rangle_c$ is 
evaluated up to the leading order which corresponds to the isotropic component, resulting in
\begin{align}
    &\left\langle P^{\rm X_1Y_1}(\bm k_1, n_1) \,P^{\rm X_2Y_2}(\bm k_2, n_2) \right\rangle_c \nonumber\\
    &= 4\pi \frac{\delta_{k_1, k_2}}{N_{k_1}(z_1)} \sum_{J_1, J_2} \mathcal{L}_{J_1}(\hat{k}_1\cdot\hat{n}_1) \mathcal{L}_{J_2}(\hat{k}_2\cdot\hat{n}_2)
    \times 4\pi \delta_{\rm D}^{(2)}(\hat{n}_1-\hat{n}_2) \nonumber\\
    & ~~~ \times \left[ \widetilde{P}^{\rm X_1X_2}_{J_1}(k_1) \widetilde{P}^{\rm Y_1Y_2}_{J_2}(k_1) \delta_{\rm D}^{(2)}(\hat{k}_1+\hat{k}_2) + \widetilde{P}^{\rm X_1Y_2}_{J_1}(k_1) \widetilde{P}^{\rm X_2Y_1}_{J_2}(k_1) \delta_{\rm D}^{(2)}(\hat{k}_1-\hat{k}_2)\right],
    \label{covariance_power}
\end{align}
where $N_{k} = 4\pi k^2\Delta kV$ is the number of available Fourier modes, with $V$
and $\Delta k$ being respectively the survey volume in each redshift slice and the width of each Fourier bin. The $\widetilde{P}^{\rm XY}_{J}$ are the Legendre coefficients of the observed power spectra including shot-noise contributions. For $\widetilde{P}^{\rm gg}_0$ and $\widetilde{P}^{\rm EE}_0$, they are respectively given by $\widetilde{P}^{\rm gg}_0 = P^{\rm gg}_0 + 1 / n_{\rm g}$, $\widetilde{P}^{\rm EE}_0 = P^{\rm EE}_0 + \sigma_\gamma^2 / n_{\rm g}$, where $n_{\rm g}$ and $\sigma_\gamma$ represent the number density of observed galaxies and the shape noise, respectively. For other cases, $\widetilde{P}^{\rm XY}_J$ is equivalent to $P_J^{\rm XY}$. 
We also assume that there is no correlation between the samples in different redshift bins.

By following the angular momentum addition procedure described in Appendix C of Ref.~\cite{Shiraishi_2017} or Appendix of Ref.~\cite{Shiraishi:2020pea}, the covariance matrix of BipoSH coefficients becomes
\begin{align}
    &\left\langle{}_{\rm X_1Y_1}\pi_{\ell_1\ell_1^\prime}^{L_1M_1}(k_1)^\ast \,_{\rm X_2Y_2}\pi_{\ell_2\ell_2^\prime}^{L_2M_2}(k_2)\right\rangle_c
    \nonumber\\
    &= \int d^2\hat{k}_1 \int d^2\hat{k}_2 \int d^2\hat{n}_1 \int d^2\hat{n}_2
    \left\langle P^{\rm X_1Y_1}(\bm k_1, \hat{n}_1) \,P^{\rm X_2Y_2}(\bm k_2, \hat{n}_2) \right\rangle_c \,
    X_{\ell_1\ell_1^\prime}^{L_1M_1}(\hat{k}_1\cdot\hat{n}_1) X_{\ell_2\ell_2^\prime}^{L_2M_2}(\hat{k}_2\cdot\hat{n}_2)^\ast\\
    &= \delta_{L,L^\prime}\delta_{M,M^\prime}\frac{\delta_{k_1, k_2}}{N_{k_1}(z_1)} \Theta_{\ell_1\ell_1^\prime \ell_2\ell_2^\prime}^{L,\,\rm X_1Y_1X_2Y_2}(k_1)
    \label{covariance_conversion},
\end{align}
where $\Theta$ is defined by
\begin{align}
    \Theta_{\ell_1\ell_1^\prime \ell_2\ell_2^\prime}^{L,\,\rm X_1Y_1X_2Y_2}(k) 
    =&~ 16\pi^2 (-1)^{\ell_1+\ell_2^\prime} \sum_{\ell,J,J^\prime} \left[\widetilde{P}_{J}^{\rm X_1X_2}(k) \widetilde{P}_{J^\prime}^{\rm Y_1Y_2}(k) +(-1)^{\ell_2}\widetilde{P}_{J}^{\rm X_1Y_2}(k) \widetilde{P}_{J^\prime}^{\rm X_2Y_1}(k)\right] \nonumber\\
    &\times (2\ell+1)\sqrt{(2\ell_1+1)(2\ell_1^\prime+1)(2\ell_2+1)(2\ell_2^\prime+1)}H_{JJ^\prime \ell}^2H_{\ell_1\ell_2\ell}H_{\ell_1^\prime \ell_2^\prime \ell}\left\{\begin{matrix}
        L&\ell_2&\ell_2^\prime\\
        \ell&\ell_1^\prime&\ell_1
    \end{matrix}\right\},
    \label{covariance_Theta}
\end{align}
with the function enclosed by the curly bracket denoting the Wigner 6$j$ symbol. 

Combining Eq.(\ref{Fisher_def})-(\ref{covariance_power}), the explicit form of the Fisher matrix is given by
\begin{align}
    F_{ij}
    = V\int_{k_{\rm min}}^{k_{\rm max}}\frac{k^2dk}{2\pi^2} \sum_{\substack{
        \ell_1, \ell_1^\prime, \ell_2, \ell_2^\prime, L, M, \\
        \rm X_1, Y_1, X_2, Y_2
    }} 
    \frac{\partial \,_{\rm X_1Y_1}\pi_{\ell_1\ell_1^\prime}^{LM}(k)^\ast}{\partial \theta_i^\ast}
    \left(\Theta^{-1}\right)_{\ell_1\ell_1^\prime \ell_2\ell_2^\prime}^{L,\rm X_1Y_1X_2Y_2}(k) \frac{\partial \,_{\rm X_2Y_2}\pi_{\ell_2\ell_2^\prime}^{LM}(k)}{\partial \theta_j} 
    ,
    \label{Fisher_explicit}
\end{align}
with the summation over wave numbers replaced by an integral.
In the following analysis, we primarily use the explicit form of the Fisher matrix given in Eq.~\eqref{Fisher_explicit}, which allows us to quantitatively evaluate the forecasted constraints on model parameters from the observed anisotropic power spectra.

In particular, the Fisher matrix is diagonal in the components corresponding to the dipolar modulation amplitude, 
$F_{A_{1M},A_{1M^\prime}}=\delta_{M,M^\prime}F_{A_{1M},A_{1M}}$.
This indicates that the constraints on each component of the amplitude of the modulation $A_{1M}$ do not depend on 
 the choice of the reference axis for defining the spherical harmonics.
\end{widetext}

Then, the $1\sigma$ error of the parameter $\theta_i$ is given through the Fisher matrix as
\begin{align}
    \Delta \theta_i = \sqrt{\left(F^{-1}\right)_{ii}}.
    \label{error_from_Fisher_single}
\end{align}
When multiple redshift bins are considered, 
where the power spectra in different bins are treated as independent,
and the Fisher matrix for each bin centered at redshift $z_n$ is denoted by $F(z_n)$,  
the total Fisher matrix is given by the sum over bins:
\begin{align}
    F_{ij}^{\mathrm{tot}} = \sum_n F_{ij}(z_n).
\end{align}
Then, the forecasted $1\sigma$ error of the parameter $\theta_i$ is given by
\begin{align}
    \Delta \theta_i = \sqrt{\left(F_{\mathrm{tot}}^{-1}\right)_{ii}}.
    \label{error_from_Fisher_combined}
\end{align}

In this paper, we consider the following two cases:  
(1). $\theta_i \in \{A_{1M}\}$, where all parameters except for the modulation amplitudes are assumed to be known.
this case is discussed in Sections \ref{Subsubsection:Fisher forecast from galaxy clustering}-\ref{Subsubsection:Dependence of IA model parameters};  
(2). $\theta_i \in \{A_{1M},\, b_1(z_j),\, b_K(z_j)\}$, where the amplitudes and the bias parameters in each redshift slice (labeled with $j$) are estimated simultaneously, discussed in Section \ref{Subsubsection:Effects of Bias Marginalization}.

\section{Results}
\label{sec:results}

\subsection{Survey Setup}

In this study, we consider two major upcoming spectroscopic galaxy surveys: Euclid \cite{Euclid2025} and DESI \cite{desicollaboration2016desi}, focusing on their Luminous Red Galaxy (LRG) and Emission Line Galaxy (ELG) samples. These surveys cover wide areas of the sky and broad redshift ranges, making them well suited for testing rotational symmetry, particularly dipolar asymmetry in the large-scale structure of the universe.

The wide sky coverage is crucial for probing large-scale anisotropies because it provides access to the largest modes in the galaxy distribution. IA is most directly established observationally for red, elliptical galaxies such as LRGs, whose shapes correlate with the large-scale tidal field \cite{Hirata_2007}. For the ELG samples, we adopt the halo-sensitive interpretation of the shape observable described in Sec.~\ref{IntrinsicAlignmentofGalaxy_IA}.

We adopt the fiducial IA amplitude $A_{\rm IA}=18$ \cite{Kurita_2020} and shape noise $\sigma_\gamma=0.3$ \cite{Lamman_2023}. As discussed in Sec.~\ref{IntrinsicAlignmentofGalaxy_IA}, $A_{\rm IA}=18$ is interpreted as a high-response benchmark for an effective halo-sensitive shape observable, rather than as the conventional central-galaxy ellipticity of all samples. Results for the more moderate value $A_{\rm IA}=5$ are presented in Appendix~\ref{app:moderate_AIA5}. The intrinsic-alignment amplitude and effective shape noise depend on the galaxy sample and the adopted estimator, and we examine their impact directly in Sec.~\ref{Subsubsection:Dependence of IA model parameters}. For the fiducial dipolar modulation amplitude $A_{1M}$, we refer to the CMB dipolar power asymmetry \cite{Aiola_2015}, whose scale-independent amplitude is approximately 0.063. We adopt the fiducial cosmological parameters from the Planck 2018 results \cite{2020_Planck_cosmological_parameter}.

The observational parameters used in our analysis are summarized in Table~\ref{Table_ExperimentalParameters}.

\begin{table}[t]
\caption{Experimental parameters adopted for the Fisher-matrix calculations in Eq.~\eqref{Fisher_explicit}. The values of $b_{\rm K}$ are determined from Eq.~\eqref{bias_and_A_IA} using the fiducial high-response value $A_{\rm IA}=18$. For an aperture-defined shape estimator, $n_{\rm g}$ is also used as the idealized shape-tracer number density.}
\label{Table_ExperimentalParameters}
    \centering
    \begin{tabular}{|c||c|c|c|c|c|}
        \hline
         & $z$ & $b_1$ & $b_{\rm K}$ & $n_{\rm g}~[h^3\,{\rm Mpc}^{-3}]$ & $V~[h^{-3}\,{\rm Gpc}^{3}]$ \\
        \hline\hline
        Euclid \cite{Euclid2025} & $1.00$ & $1.46$ & $-0.123$ & $6.86\times 10^{-4}$ & $7.94$\\
         & $1.20$ & $1.61$ & $-0.134$ & $5.58\times 10^{-4}$ & $9.15$\\
         & $1.40$ & $1.75$ & $-0.145$ & $4.21\times 10^{-4}$ & $10.05$\\
         & $1.65$ & $1.90$ & $-0.159$ & $2.61\times 10^{-4}$ & $16.22$\\
        \hline
        DESI LRG & $0.65$ & $2.34$ & $-0.105$  & $4.4\times 10^{-4}$ & $2.63$\\
        \cite{desicollaboration2016desi} & $0.75$ & $2.49$ & $-0.110$ & $4.2\times 10^{-4}$ & $3.15$\\
         & $0.85$ & $2.61$ & $-0.115$ & $2.5\times 10^{-4}$ & $3.65$\\
         & $0.95$ & $2.73$ & $-0.121$ & $9.3\times 10^{-5}$ & $4.10$\\
         & $1.05$ & $2.86$ & $-0.126$ & $1.6\times 10^{-5}$ & $4.52$\\
         & $1.15$ & $2.98$ & $-0.131$ & $4.9\times 10^{-6}$ & $4.89$\\
        \hline
        DESI ELG & $0.65$ & $1.17$ & $-0.105$ & $1.6 \times 10^{-4}$ & $2.63$ \\
        \cite{desicollaboration2016desi} & $0.75$ & $1.23$ & $-0.110$ & $1.0 \times 10^{-3}$ & $3.15$ \\
         & $0.85$ & $1.29$ & $-0.115$ & $7.4 \times 10^{-4}$ & $3.65$ \\
         & $0.95$ & $1.35$ & $-0.121$ & $7.2 \times 10^{-4}$ & $4.10$ \\
         & $1.05$ & $1.41$ & $-0.126$ & $4.5 \times 10^{-4}$ & $4.52$ \\
         & $1.15$ & $1.47$ & $-0.131$ & $3.9 \times 10^{-4}$ & $4.89$ \\
         & $1.25$ & $1.53$ & $-0.137$ & $3.6 \times 10^{-4}$ & $5.22$ \\
         & $1.35$ & $1.60$ & $-0.142$ & $1.3 \times 10^{-4}$ & $5.50$ \\
         & $1.45$ & $1.66$ & $-0.148$ & $1.1 \times 10^{-4}$ & $5.75$ \\
         & $1.55$ & $1.72$ & $-0.154$ & $7.7 \times 10^{-5}$ & $5.97$ \\
         & $1.65$ & $1.78$ & $-0.159$ & $2.9 \times 10^{-5}$ & $6.15$ \\
        \hline
\end{tabular}
\end{table}

\subsection{Signal-to-noise ratio}
Before presenting the Fisher analysis, we examine the signal-to-noise ratio (SNR), defined as
\begin{widetext}
\begin{align}
    {\rm SNR}^2
    &= \sum_{k_1,k_2}^{k_{\rm max}}\sum_{\mathcal{C}_1\mathcal{C}_2}
    \;_{\rm X_1Y_1}\pi_{\ell_1\ell_1^\prime}^{L_1M_1}(k_1)^\ast
    \left\langle\;_{\rm X_1Y_1}\pi_{\ell_1\ell_1^\prime}^{L_1M_1}(k_1)^\ast \;_{\rm X_2Y_2}\pi_{\ell_2\ell_2^\prime}^{L_2M_2}(k_2)\right\rangle_c^{-1}
    \;_{\rm X_2Y_2}\pi_{\ell_2\ell_2^\prime}^{L_2M_2}(k_2)
    \nonumber\\
    &= V\int_{k_{\rm min}}^{k_{\rm max}}\frac{k^2dk}{2\pi^2} \sum_{\substack{
        \ell_1, \ell_1^\prime, \ell_2, \ell_2^\prime, L, M, \\
        \rm X_1, Y_1, X_2, Y_2}}
        \;_{\rm X_1Y_1}\pi_{\ell_1\ell_1^\prime}^{LM}(k)^\ast\, \left(\Theta^{-1}\right)_{\ell_1\ell_1^\prime \ell_2\ell_2^\prime}^{L,\rm X_1Y_1X_2Y_2}(k)
        \;_{\rm X_2Y_2}\pi_{\ell_2\ell_2^\prime}^{LM}(k)
    \label{SNR_def_and_explicit}.
\end{align}
\end{widetext}
When combining multiple redshift bins centered at $z_n$, we sum their squared SNRs,
\begin{align}
    \mathrm{SNR}_{\mathrm{tot}}^2 = \sum_n \mathrm{SNR}^2(z_n).
\end{align}
This is analogous to summing the Fisher matrices over redshift bins.

\begin{figure*}[t]
\includegraphics[scale=0.42]{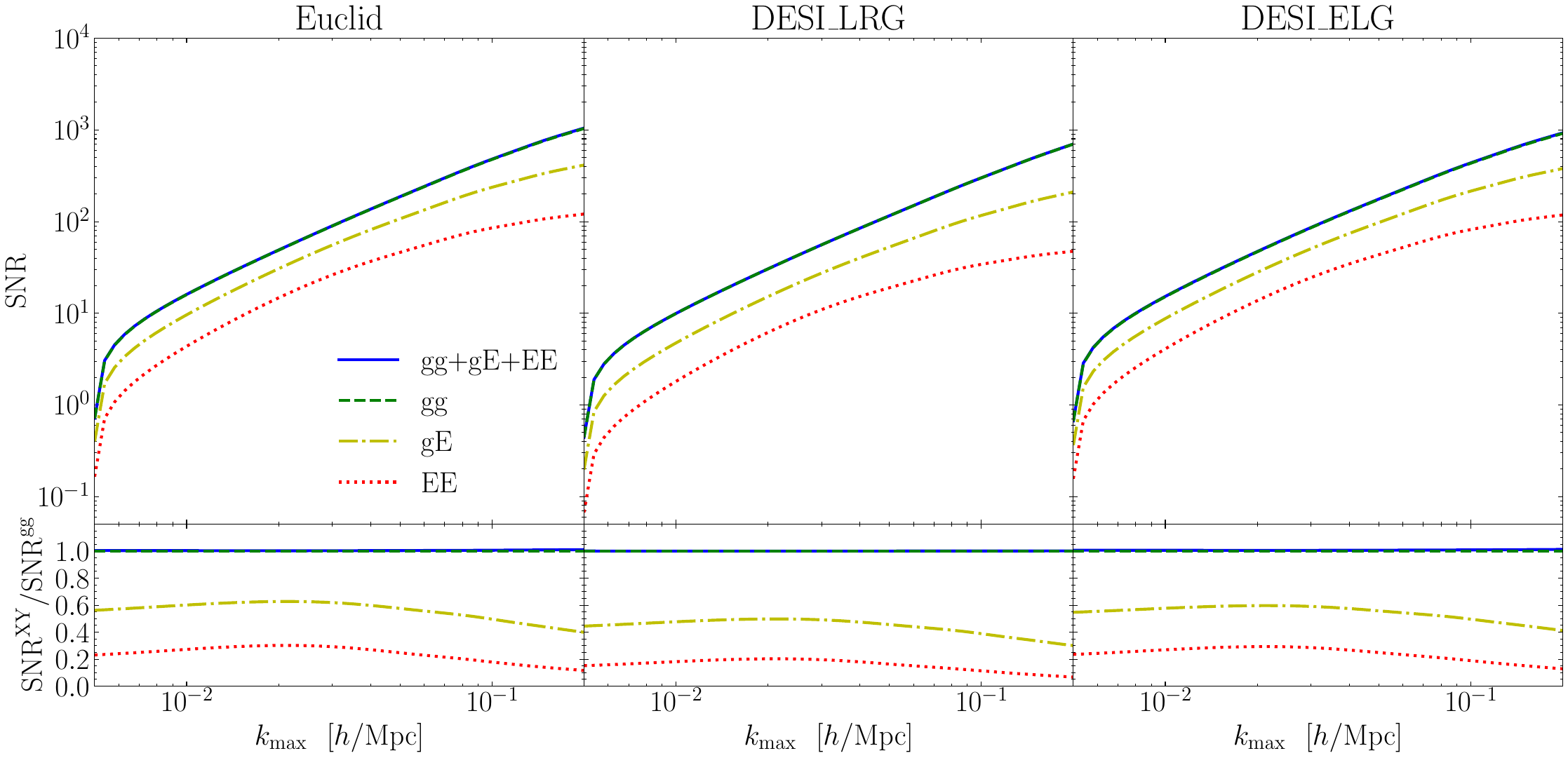}
\caption{{\it Top panels}: cumulative SNRs as functions of $k_{\rm max}$ for each survey, using the fiducial value $A_{\rm IA}=18$. The green, yellow, and red curves represent the individual SNRs of $\,_{\rm gg}\pi_{\ell\ell^\prime}^{LM}$, $\,_{\rm gE}\pi_{\ell\ell^\prime}^{LM}$, and $\,_{\rm EE}\pi_{\ell\ell^\prime}^{LM}$, respectively, while the blue curve represents the joint SNR. {\it Bottom panels}: SNRs normalized by that obtained from $P^{\rm gg}$ alone.\label{Fig_SNR}}
\end{figure*}

Unless otherwise stated, the main-text results use the fiducial high-response value $A_{\rm IA}=18$. Figure~\ref{Fig_SNR} shows the cumulative SNR as a function of $k_{\rm max}$ for the individual spectra $P^{\rm gg}$, $P^{\rm gE}$, and $P^{\rm EE}$ and for their joint analysis. The total SNR is not substantially increased by adding IA to galaxy clustering. Nevertheless, in some survey configurations the cross spectrum $P^{\rm gE}$ contains nearly half as much information as $P^{\rm gg}$. This makes it possible to compare two observationally distinct estimates of the dipolar modulation, one from galaxy clustering and the other from the density--shape cross correlation.

\begin{figure*}[t]
\includegraphics[scale=0.42]{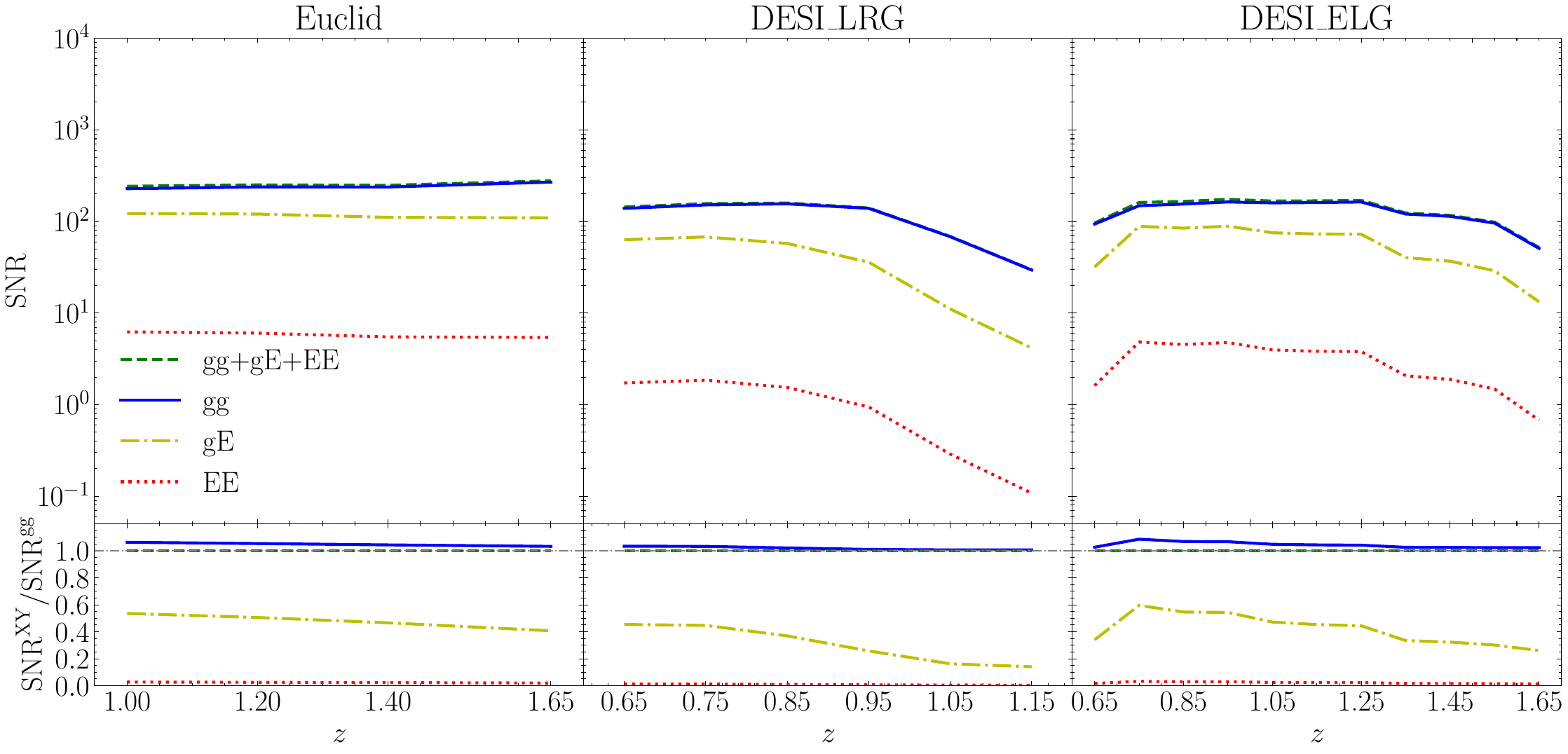}
\caption{{\it Top panels}: SNRs in each redshift bin for each survey, using $A_{\rm IA}=18$. {\it Bottom panels}: SNRs normalized by that obtained from $P^{\rm gg}$ in the same bin. The analysis is performed with $k_{\rm max}=0.1\,h\,{\rm Mpc}^{-1}$.\label{Fig_SNR_redshift}}
\end{figure*}

Figure~\ref{Fig_SNR_redshift} shows the SNR contribution from each redshift bin, evaluated with the fixed scale cut $k_{\rm max}=0.1\,h\,{\rm Mpc}^{-1}$. For Euclid, the contribution is broadly distributed across the full redshift range. For DESI, the dominant contribution comes from $z<1.0$ in the LRG sample and from $z\simeq0.75$--$1.25$ in the ELG sample, corresponding to bins with relatively high galaxy number densities. The IA-related results should be interpreted under the aperture-shape assumption described above.


\subsection{Constraints on Dipolar Modulation Amplitude}

We next present the forecast constraints on the dipolar modulation amplitude $A_{1M}$ for the fiducial value $A_{\rm IA}=18$. The corresponding results for $A_{\rm IA}=5$ are shown in Appendix~\ref{app:moderate_AIA5}.

\subsubsection{Fisher forecast from galaxy clustering}\label{Subsubsection:Fisher forecast from galaxy clustering}

\begin{figure*}[t]
\includegraphics[scale=0.42]{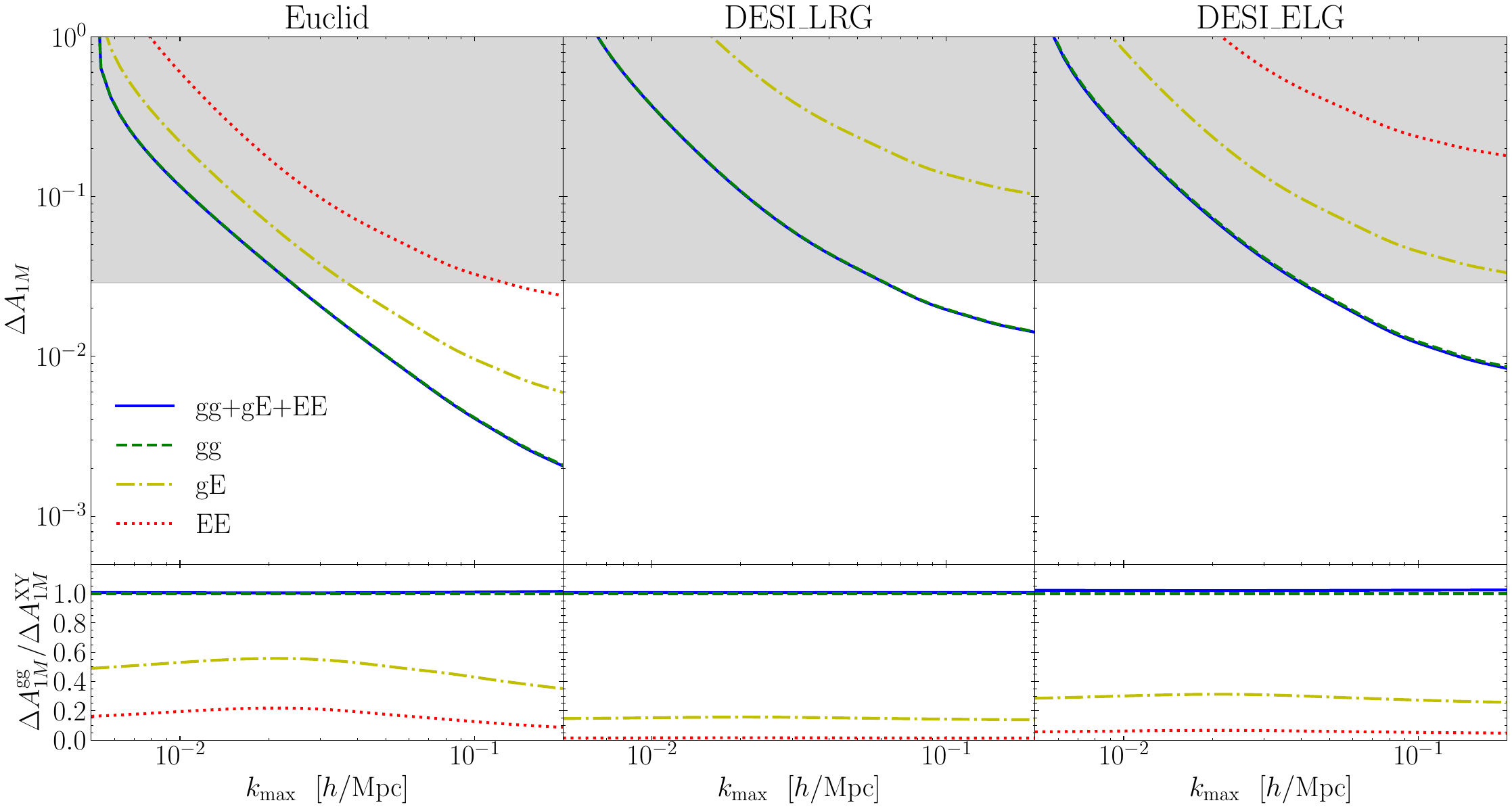}
\caption{{\it Top panels}: expected $1\sigma$ errors as functions of $k_{\rm max}$ for each survey, assuming $f_{\rm mod}(k)=1$ and $A_{\rm IA}=18$. The region where the error is larger than the CMB constraint \cite{Aiola_2015} is shaded in gray. Although three different $M$ modes are considered, the constraints do not depend on $M$. {\it Bottom panels}: errors normalized by the error obtained from $P^{\rm gg}$. The color scheme is the same as in Fig.~\ref{Fig_SNR}.\label{Fig_error_const}}
\end{figure*}

\begin{figure*}
    \centering
    \includegraphics[scale=0.42]{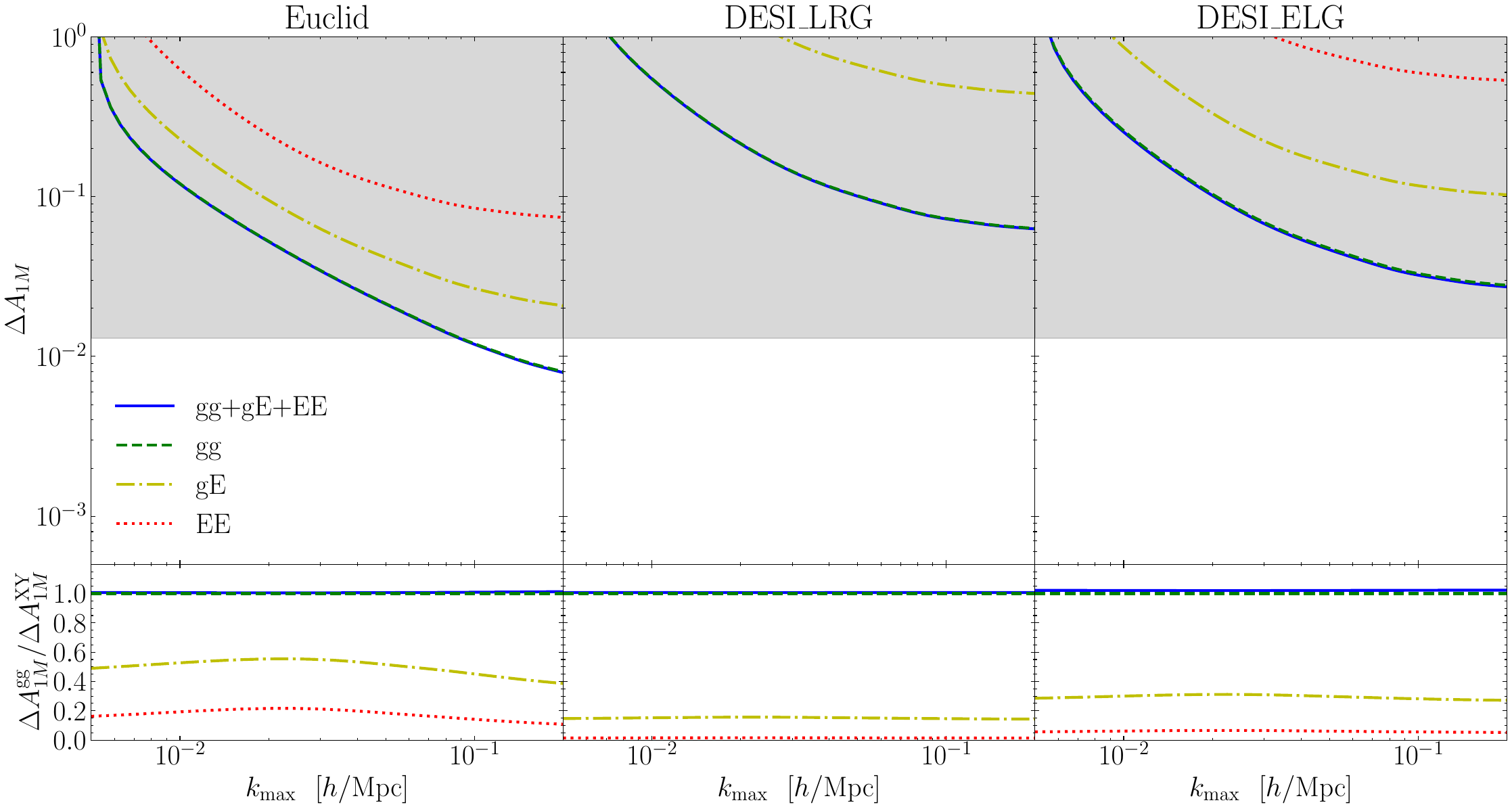}
    \caption{Same as Fig.~\ref{Fig_error_const}, but for $f_{\rm mod}(k)=(k/k^{\rm c})^{-0.5}$ with $k^{\rm c}=0.005\,{\rm Mpc}^{-1}$.}
    \label{Fig_error_05}
\end{figure*}

We first perform a Fisher analysis using galaxy clustering alone. This provides the baseline against which the IA contribution is assessed. The top panels of Fig.~\ref{Fig_error_const} show the forecast $1\sigma$ errors on $A_{1M}$ for a scale-independent modulation, $f_{\rm mod}(k)=1$, with fiducial value $A_{1M}=0.063$ \cite{Aiola_2015}. The green curves show the errors from $P^{\rm gg}$ alone. The gray region indicates errors larger than the CMB constraint of Ref.~\cite{Aiola_2015}. Figure~\ref{Fig_error_05} shows the corresponding results for $f_{\rm mod}(k)=(k/k^{\rm c})^{-0.5}$, where $k^{\rm c}=0.005\,{\rm Mpc}^{-1}$ and the fiducial value is $A_{1M}=0.03$ \cite{Aiola_2015}. The scale dependence suppresses the signal at small scales and therefore weakens the constraint relative to the scale-independent case.

The results show that, depending on the survey configuration and the scale dependence of the modulation, galaxy clustering alone can surpass the current CMB constraint. Our result is consistent with the forecast of Ref.~\cite{Shiraishi_2017}, while updating the survey specifications and redshift binning.

\subsubsection{Joint analysis with Intrinsic Alignments}\label{Subsubsection:Joint analysis with Intrinsic Alignments}

The yellow and red curves in Figs.~\ref{Fig_error_const} and \ref{Fig_error_05} show the errors obtained from $P^{\rm gE}$ and $P^{\rm EE}$, respectively, while the blue curves show the joint constraint from all spectra. The inclusion of IA does not drastically improve the combined constraint because $P^{\rm gg}$ already dominates the total information. However, for the fiducial high-response value, $P^{\rm gE}$ can provide a constraint less than twice as weak as that from $P^{\rm gg}$ in favorable cases. Thus the density--shape cross spectrum retains non-negligible constraining power even when its addition produces only a modest change in the joint statistical error.

The signal-to-noise ratio is not generally identical to Fisher information. In the present model, however, the modulation amplitude $A_{1M}$ enters $P^{\rm gg}$ and $P^{\rm gE}$ with the same scale and angular dependence. Their relative SNRs therefore closely trace their relative information on $A_{1M}$, consistently with the Fisher results. Since the dominant systematics of galaxy clustering and shape measurements differ, agreement between the estimates from $P^{\rm gg}$ and $P^{\rm gE}$ would provide a useful test against survey-window, imaging, and data-processing systematics.

\subsubsection{Dependence of IA model parameters}
\label{Subsubsection:Dependence of IA model parameters}

\begin{figure*}
    \centering
    \includegraphics[scale=0.42]{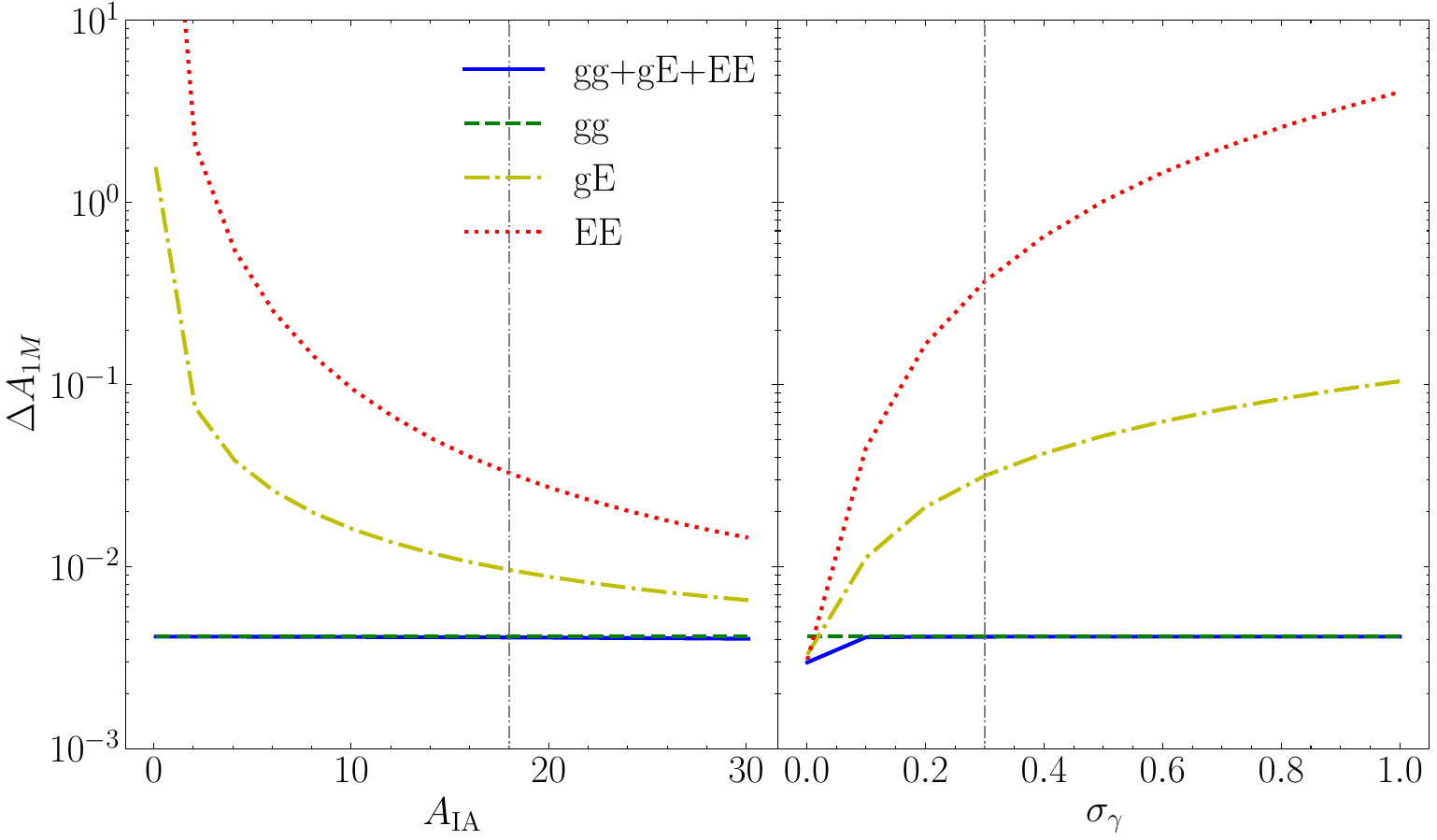}
    \caption{
    Expected $1\sigma$ errors for Euclid with varying parameters:
    In the left panel, $A_{\rm IA}$ is varied, while in the right
    panel, $\sigma_\gamma$ is varied.
    The dot-dashed lines indicate the fiducial values
    $A_{\rm IA}=18$ and $\sigma_\gamma=0.3$.
    The analysis is performed with
    $k_{\rm max}=0.1\,h\mathrm{Mpc}^{-1}$.
    }
    \label{Fig_error_variating_parameters}
\end{figure*}

We also evaluated Eq.~(\ref{Fisher_explicit}) while varying
$A_{\rm IA}$ and $\sigma_\gamma$ from their fiducial values.
The results are shown in Fig.~\ref{Fig_error_variating_parameters}
, using the fixed scale cut
$k_{\rm max}=0.1\,h\mathrm{Mpc}^{-1}$.
We focus on Euclid because its relatively low galaxy bias and high
number density give the largest relative contribution from IA among
the survey specifications considered here.

We find that the constraint from IA alone is strongly dependent on
both $A_{\rm IA}$ and $\sigma_\gamma$, as these parameters directly
control the amplitude and noise level of the IA signal, respectively.
In particular, smaller values of $\sigma_\gamma$ improve the
sensitivity to the IA signal, thereby enhancing the constraining
power.

Among the IA observables, the cross spectrum $P^{\rm gE}$ provides
the dominant constraining power. Importantly, over a broad range of
nonzero values of $A_{\rm IA}$ considered here, $P^{\rm gE}$ alone
yields a meaningful standalone constraint on the dipolar modulation
amplitude. Even for a moderate response of $A_{\rm IA}\sim5$, the
constraint from $P^{\rm gE}$ can be comparable to
the current CMB constraint of Ref.~\cite{Aiola_2015}. For larger
values of $A_{\rm IA}$, $P^{\rm gE}$ provides a substantial fraction
of the constraining power of $P^{\rm gg}$.

On the other hand, the constraint obtained by combining IA with
galaxy clustering shows only mild dependence on these parameters,
indicating that the contribution from IA is subdominant
in the combined statistical error. Thus, the main utility of
$P^{\rm gE}$ is not necessarily a large reduction in the final
combined uncertainty. Rather, it provides an observationally
distinct determination of the modulation amplitude over a wide
range of possible IA responses and can therefore serve as a
consistency check of the constraint inferred from $P^{\rm gg}$.
Detailed results for the case $A_{\rm IA}=5$ are presented in
Appendix~\ref{app:moderate_AIA5}.

\subsubsection{Effects of Bias Marginalization}\label{Subsubsection:Effects of Bias Marginalization}

Next, we consider a more practical scenario that includes the marginalization over other parameters. 
In the previous section and previous studies, the analysis was performed under the assumption that only the amplitude of the dipolar modulation, $A_{1M}$, was estimated simultaneously. 
However, in practical analyses, it is necessary to consider the case where other parameters, such as the bias parameters, $b_1$ and $b_{\rm K}$, are also simultaneously estimated. 

We first analytically find that the change in accuracy due to the marginalization is of the order of $A_{1M}^2$, because the off-diagonal terms of the Fisher matrix are of the order of $A_{1M}$, 
as detailed in Appendix \ref{effects_of_marginalizing_over_bias_parameters} (see Eq.\,\eqref{OffDiagonal_A1M}).
This result implies that as long as $A_{1M}$ remains small, marginalizing over other parameters has little impact on the constraint of $A_{1M}$ itself. 
In other words, the effect of marginalization becomes significant only when $A_{1M}$ grows to the order of unity.

We then numerically examine this effect by considering a specific scale-dependent modulation model with $f_{\rm mod}(k) = (k/k^c)^2$. 
This model is motivated by the desire to investigate how the impact of marginalizing over bias parameters depends on the modulation amplitude, and also enables us to explore the regime where $A_{1M} \ll 1$ no longer holds.

In this analysis, we treat the redshift-dependent bias parameters $b_1(z)$ and $b_{\rm K}(z)$ as independent parameters in each redshift bin, and marginalize over them separately. 
This allows us to evaluate how the marginalization of redshift-dependent bias parameters affects the constraints on the modulation amplitude.

\begin{figure*}
    \centering
    \includegraphics[scale=0.42]{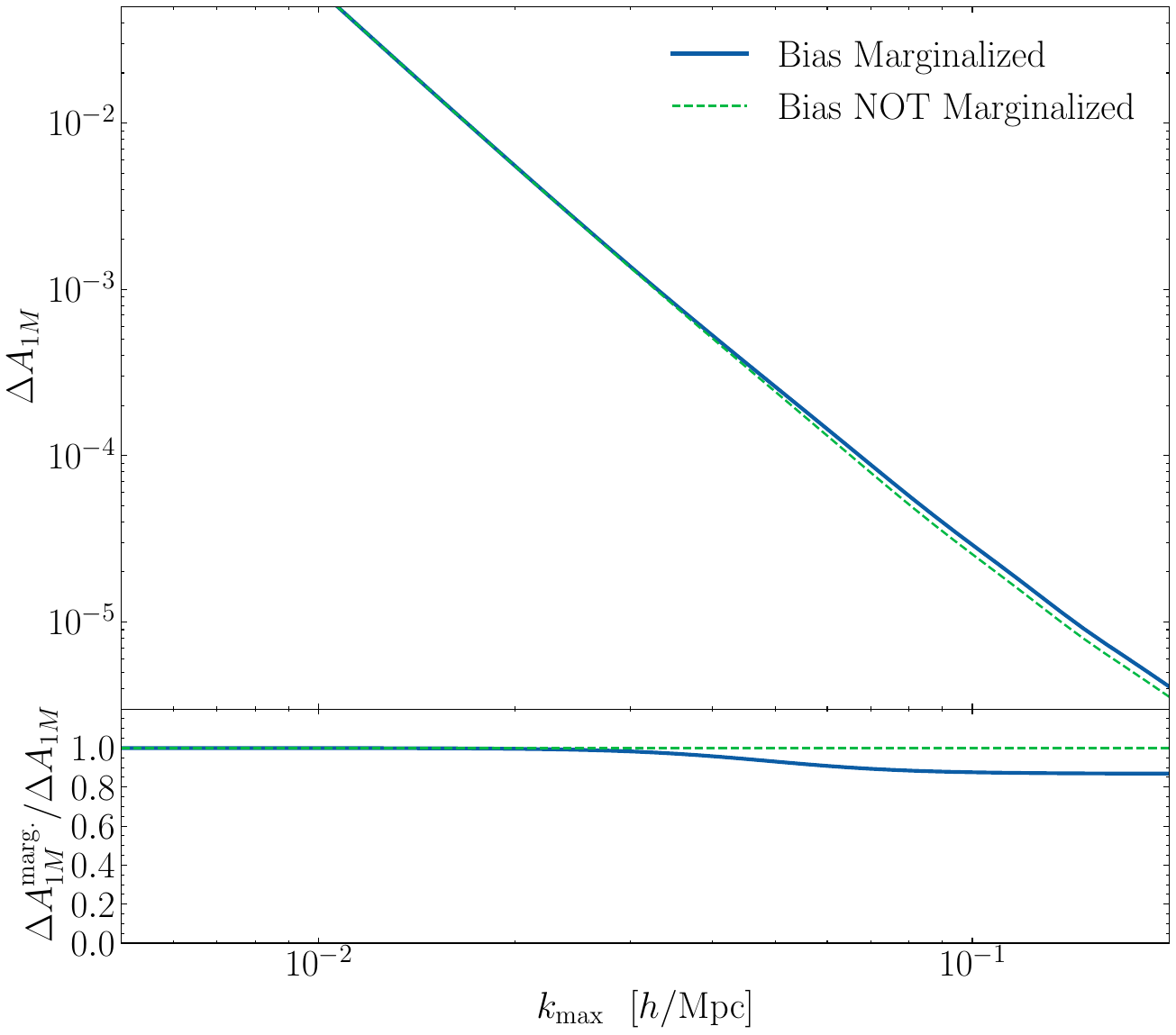}
    \caption{
    {\it Top panel}: expected $1\sigma$ errors for $f_{\rm mod}(k) = \left({k}/{k^{\rm c}}\right)^{2}$. The blue (green) curve shows the expected $1\sigma$ error with (without) marginalizing over the bias parameters $b_1(z)$ and $b_{\rm K}(z)$. 
    {\it Bottom panel}: expected errors with marginalizing over the bias parameters, based on the error without marginalization.
    The result of the forecast combining IA with clustering is shown in this figure.
    Note that the result changes very little even when using galaxy clustering alone.
    }
    \label{Fig_error_2}
\end{figure*}

The result is shown in Fig.~\ref{Fig_error_2}, which indicates that there is only a weak degeneracy between the modulation amplitude $A_{1M}$ and the bias parameters $b_1(z)$ and $b_{\rm K}(z)$, particularly in the region $k\lesssim0.03~h/{\rm Mpc}$, where $A_{1M}$ is small. 
The figure shows the result of the forecast combining IA with galaxy clustering, but we have confirmed that the result changes very little even when using galaxy clustering alone.
In the regime $k\gtrsim0.03~h/{\rm Mpc}$ where $A_{1M} \gtrsim 1$, the off-diagonal elements of the Fisher matrix become non-negligible, and the marginalization over bias parameters leads to a noticeable degradation in the constraint accuracy.

\section{Conclusion}
\label{sec:conclusion}
In this study, we investigated a dipolar power asymmetry as a possible signature of statistical anisotropy in primordial fluctuations. Building on the BipoSH analysis of Ref.~\cite{Shiraishi_2017}, we incorporated intrinsic-alignment auto and cross correlations into a Fisher forecast for Euclid and DESI survey specifications.

We found that IA only modestly improves the combined constraint on the dipolar modulation because galaxy clustering dominates the total Fisher information. Nevertheless, in favorable high-number-density samples, the density--shape cross spectrum $P^{\rm gE}$ can carry up to about half the constraining power of $P^{\rm gg}$. Its principal value is therefore as an independent consistency check: a signal recovered from both galaxy clustering and a halo-sensitive shape observable would be less likely to arise from a systematic affecting only one measurement.

We also examined the dependence on $A_{\rm IA}$ and $\sigma_\gamma$. IA-only constraints vary strongly with these parameters, while the joint constraint is comparatively stable. The moderate case $A_{\rm IA}=5$ weakens the standalone IA constraints but does not alter the qualitative conclusion that the cross spectrum supplies complementary information rather than a dramatic reduction of the combined statistical error.

Finally, we studied marginalization over the galaxy bias and IA bias parameters. Analytically and numerically, its effect on the modulation constraint is negligible when the modulation amplitude is small, because the relevant off-diagonal Fisher-matrix elements are proportional to $A_{1M}$. The BipoSH framework combined with galaxy clustering and a calibrated shape estimator therefore provides a robust route to testing large-scale rotational symmetry with future spectroscopic and imaging data.

\begin{acknowledgments}
   We are grateful to Toshiki Kurita for his valuable insights and discussions
throughout the early stages of this project. This work was supported in part
by JST SPRING Grant No. JPMJSP2110 (K. M.), JSPS KAKENHI Grants
No. 23K20844, No. 23K25868, and No. 26H02044 (A. T.), and
No. JP20H05859 and No. JP23K03390 (M. S.). A. T. acknowledges support
from FY2025 MUSUBIME of Kyoto University. T. O. acknowledges support
from the Taiwan National Science and Technology Council under Grants
No. NSTC 112-2112-M-001-034-, No. NSTC 113-2112-M-001-011-, and
No. NSTC 114-2112-M-001-004-, and the Academia Sinica Investigator
Project Grant No. AS-IV-114-M03 for the period of 2025–2029.
M. S. acknowledges the Center for Computational Astrophysics,
National Astronomical Observatory of Japan, for providing the computing
resources.
\end{acknowledgments}

\appendix

\section{Legendre coefficients of power spectra}\label{Legendre coefficients of power spectra}
In this appendix, we show the explicit form of the Legendre coefficients $P_\ell^{\rm XY}(k)$ defined in Eq.(\ref{power_Legendre_decomposition}).
From Eq.(\ref{gg_power_explicit})-(\ref{EE_power_explicit}), they are evaluated as follows:
\begin{align}
    &P^{\rm gg}_0(k) = \left(b_1^2 + \frac{2}{3}b_1f + \frac{1}{5}f^2\right)\Bar{P}_{\rm m}(k),\\
    &P^{\rm gg}_2(k) = \left(\frac{4}{3}b_1f + \frac{4}{7}f^2\right)\Bar{P}_{\rm m}(k),\\
    &P^{\rm gg}_4(k) = \frac{8}{35}f^2\Bar{P}_{\rm m}(k),\\
    &P^{\rm gg}_1(k) = P^{\rm gg}_3(k) = P^{\rm gg}_{l\geq 5}(k) = 0,\\
    \nonumber\\
    &P^{\rm gE}_0(k) = b_{\rm K}\left(\frac{2}{3}b_1 + \frac{2}{15}f\right)\Bar{P}_{\rm m}(k),\\
    &P^{\rm gE}_2(k) = b_{\rm K}\left(-\frac{2}{3}b_1 + \frac{2}{21}f\right)\Bar{P}_{\rm m}(k),\\
    &P^{\rm gE}_4(k) = \frac{8}{35}b_{\rm K}f\Bar{P}_{\rm m}(k),\\
    &P^{\rm gE}_1(k) = P^{\rm gE}_3(k) = P^{\rm gE}_{l\geq 5}(k) = 0,\\
    \nonumber\\
    &P^{\rm EE}_0(k) = \frac{8}{15}b_{\rm K}^2\Bar{P}_{\rm m}(k),\\
    &P^{\rm EE}_2(k) = -\frac{16}{21}b_{\rm K}^2\Bar{P}_{\rm m}(k),\\
    &P^{\rm EE}_4(k) = \frac{8}{35}b_{\rm K}^2\Bar{P}_{\rm m}(k),\\
    &P^{\rm EE}_1(k) = P^{\rm EE}_3(k) = P^{\rm EE}_{l\geq 5}(k) = 0.
\end{align}

\section{Comparison with a moderate IA amplitude}
\label{app:moderate_AIA5}

The main text adopts $A_{\rm IA}=18$ as the fiducial high-response benchmark. Here we show the corresponding results for a more moderate value, $A_{\rm IA}=5$.

\begin{figure*}[t]
\includegraphics[scale=0.42]{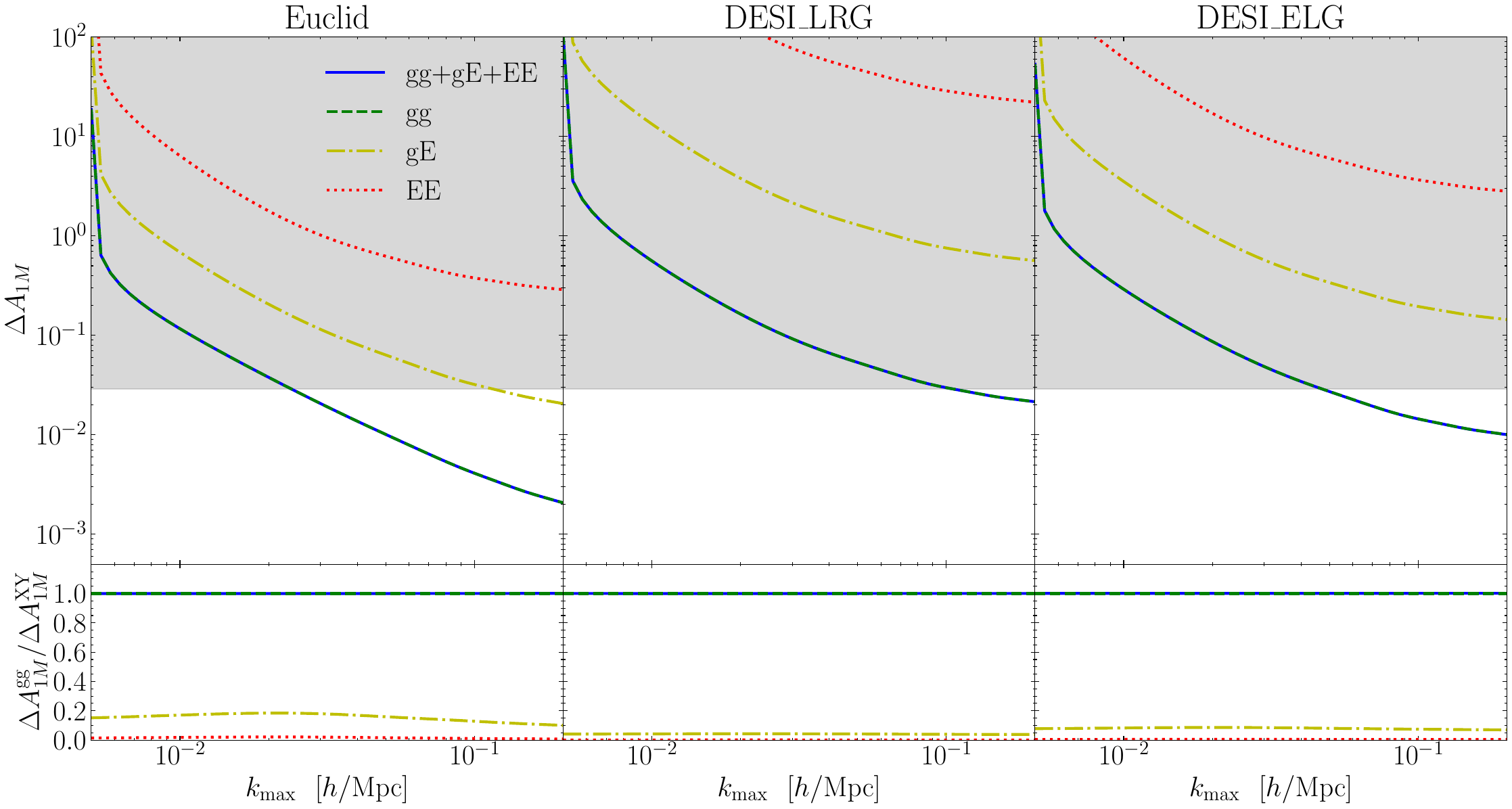}
\caption{Same as Fig.~\ref{Fig_error_const}, but for the moderate IA amplitude $A_{\rm IA}=5$. The horizontal axis varies $k_{\rm max}$.}
\label{Fig_error_const_AIA5}
\end{figure*}

\begin{figure*}[t]
\centering
\includegraphics[scale=0.42]{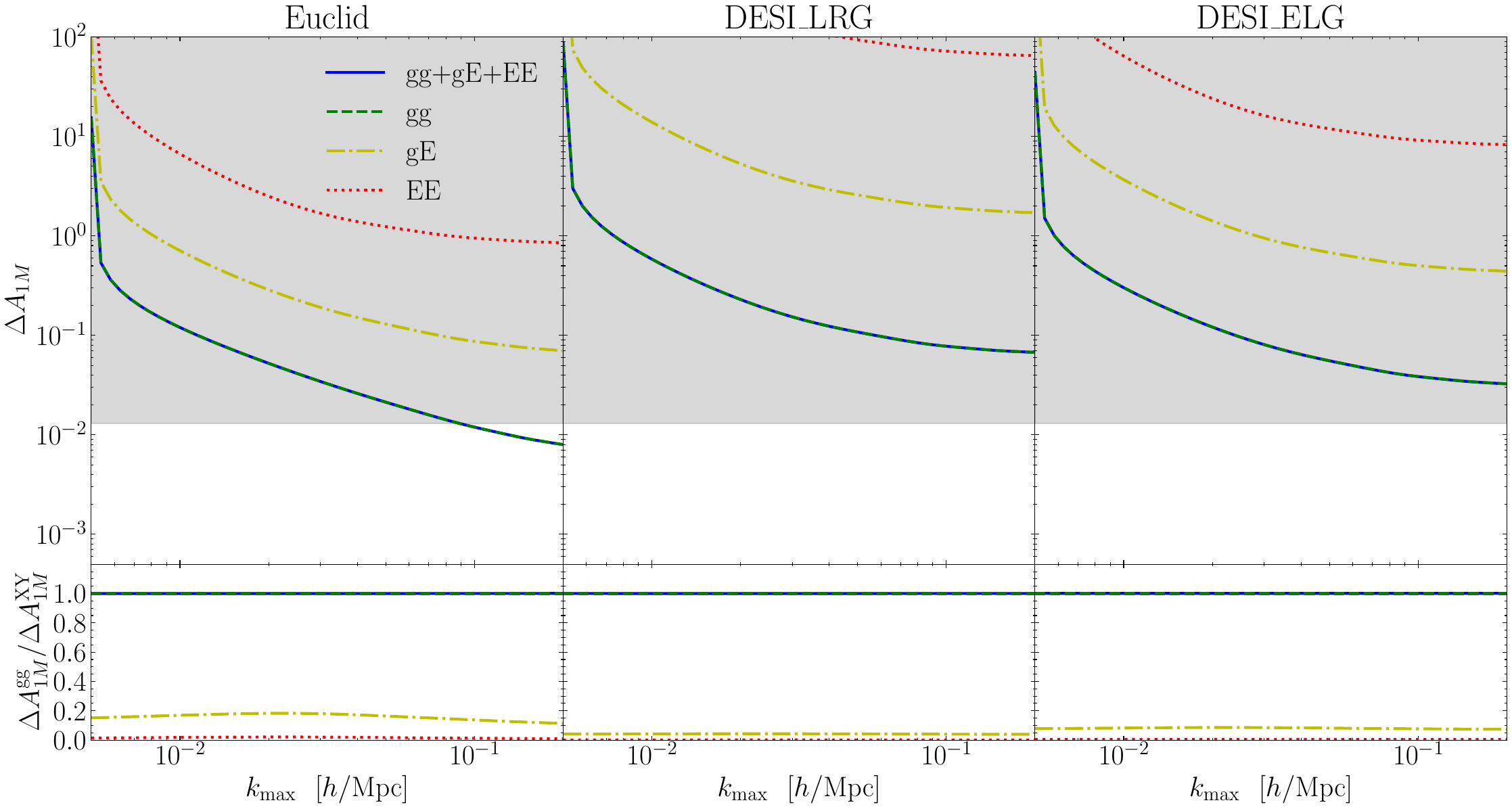}
\caption{Same as Fig.~\ref{Fig_error_05}, but for the moderate IA amplitude $A_{\rm IA}=5$. The horizontal axis varies $k_{\rm max}$.}
\label{Fig_error_05_AIA5}
\end{figure*}

The Fisher constraints for the scale-independent and scale-dependent modulation models are shown in Figs.~\ref{Fig_error_const_AIA5} and \ref{Fig_error_05_AIA5}, respectively. Reducing $A_{\rm IA}$ weakens the constraints from $P^{\rm gE}$ and $P^{\rm EE}$, while the joint constraint changes only modestly because it is dominated by $P^{\rm gg}$. Nevertheless, for the scale-independent modulation model, the Euclid constraint from $P^{\rm gE}$ alone remains comparable to the current CMB constraint. This illustrates that density--shape correlations can still provide useful information even for a more moderate IA response. On the other hand, the strong dependence of the IA-only constraints on $A_{\rm IA}$, discussed in Sec.~\ref{Subsubsection:Dependence of IA model parameters} and shown in Fig.~\ref{Fig_error_variating_parameters}, indicates that substantially smaller values would rapidly diminish their constraining power.

\section{Non-degeneracy Between Anisotropy and Biases} \label{effects_of_marginalizing_over_bias_parameters}
In this appendix, we show that the off-diagonal elements of the Fisher matrix are the order of $A_{1M}$.

Considering a symmetry matrix
\begin{align}
    M = \left(\begin{matrix}
        a&c\\
        c&b
    \end{matrix}\right),
\end{align}
with conditions $a,\,b = \mathcal{O}(1)$ and $c\ll 1$, the inverse of $M$ can be written as
\begin{align}
    M^{-1} &=\frac{1}{ab-c^2}\left(\begin{matrix}
        b&-c\\
        -c&a
    \end{matrix}\right)\nonumber\\
    &\sim \frac{1}{ab}\left(1+\frac{c^2}{ab}\right) \left(\begin{matrix}
        b&-c\\
        -c&a
    \end{matrix}\right)\nonumber\\
    &= \left(\begin{matrix}
        \frac{1}{a}\left(1+\frac{c^2}{ab}\right)&-\frac{c}{ab}\left(1+\frac{c^2}{ab}\right)\\
        -\frac{c}{ab}\left(1+\frac{c^2}{ab}\right)&\frac{1}{b}\left(1+\frac{c^2}{ab}\right)
    \end{matrix}\right).
\end{align}
Therefore, the difference between $(M^{-1})_{11}$ and $(M_{11})^{-1}$ becomes $\mathcal{O}(c^2)$.

We now apply this analysis to the Fisher matrix $F_{ij}$ with indices $i,j \in \{A_{1M},\,b_1\}$. Its diagonal elements $F_{A_{1M},A_{1M}}$ and $F_{b_1,b_1}$ are $\mathcal{O}(1)$, so the difference of $\Delta A_{1M}$ with marginalization over bias parameters and that without marginalization is the order of $A_{1M}^2$ when $A_{1M}$ is small enough and the off-diagonal element of the Fisher matrix $F_{A_{1M},b_1}$ is the order of $A_{1M}$.
$F_{A_{1M}, b_1} = \mathcal{O}(A_{1M})$ is derived as follows.

Since the covariance matrix of BipoSH coefficient ${\bf Cov}_{LL^\prime}$, which is explicitly given in Eq.\eqref{covariance_conversion},
is proportional to $\delta_{L,L^\prime}$
(see Eq.\eqref{covariance_conversion}), 
different $L$ modes are uncorrelated, and the covariance matrix becomes 
\begin{align}
    {\bf Cov}_{LL^\prime} = 
    \delta_{L,L^\prime}\bm C_L
\end{align}
where $\bm C_L$ is covariance matrix calculated from the BipoSH coefficient $_{\rm XY}\pi^{LM}_{\ell\ell^\prime}$
.

Then, Fisher matrix becomes
\begin{align}
    F_{ij} &= \sum \frac{\partial\bm \pi^{L \ast}}{\partial\theta_i^\ast}
    \left({\bf Cov}^{-1} \right)_{LL^\prime}
    \left(\frac{\partial\bm \pi^{L^\prime}}{\partial\theta_j} \right)^\top
    \nonumber\\
    &= \sum 
    \frac{\partial\bm \pi^{L \ast}}{\partial\theta_i^\ast}
    \bm C_L^{-1}
    \left(\frac{\partial\bm \pi^{L}}{\partial\theta_j}\right)^\top
\end{align}
with $\bm \pi^{L}$ representing $_{\rm XY}\pi^{LM}_{\ell\ell^\prime}$.

From Eq.\eqref{BipoSH_coefficient_explicit}, the BipoSH coefficient $\boldsymbol{\pi}^L$ is
\begin{align}
    \boldsymbol{\pi}^L = \boldsymbol{\mathcal{O}}(1)\delta_{L,0}
    +\boldsymbol{\mathcal{O}}(A_{1M})\delta_{L,1}
    ,
\end{align}
so
the signal part of Fisher matrix is
\begin{align}
    &\frac{\partial\bm \pi^{L}}{\partial A_{1M}}
    = \boldsymbol{\mathcal{O}}(1) \delta_{L,1}
    ,\\
    &\frac{\partial\bm \pi^{L}}{\partial b_1}
    = \boldsymbol{\mathcal{O}}(1) \delta_{L,0}+\boldsymbol{\mathcal{O}}(A_{1M})\delta_{L,1}
    .
\end{align}
Then the off-diagonal element of Fisher matrix $F_{A_{1M},b_1}$ is evaluated as
\begin{align}
    F_{A_{1M},b_1} = \mathcal{O}({A_{1M}})
    \label{OffDiagonal_A1M}
\end{align}.

\bibliography{apssamp}

\end{document}